\documentclass[journal]{IEEEtran}

\ifCLASSINFOpdf
\else
\fi
\usepackage{multirow}
\usepackage{amsthm}
\usepackage{graphicx}
\usepackage{color}
\usepackage{epstopdf}
\usepackage{amsfonts}
\usepackage[caption=false, font=footnotesize]{subfig}
\usepackage{cite}
\usepackage{hyperref}
\usepackage{amssymb}
\usepackage{amsmath}
\usepackage{algpseudocode,algorithm}
\usepackage{setspace}
\graphicspath{{Figures/}}
\newtheorem{lemma}{Lemma}
\newtheorem{remark}{Remark}

\usepackage{svg}
\usepackage{xcolor, soul}
  \usepackage{subfig}
  \DeclareCaptionLabelSeparator{periodspace}{.\quad}
 \usepackage{algpseudocode}  

\graphicspath{{Figures/}}
\DeclareMathSizes{10}{9}{7}{6}
\IEEEaftertitletext{\vspace{-3\baselineskip}}

\begin{document}
\title {HARQ Optimization for Real-Time Remote
Estimation in Wireless Networked Control}  
\author{\IEEEauthorblockN{Faisal Nadeem,~\IEEEmembership{Member,~IEEE}, Yonghui Li,~\IEEEmembership{Fellow,~IEEE}, 
Branka Vucetic,~\IEEEmembership{Life Fellow,~IEEE}, Mahyar Shirvanimoghaddam,~\IEEEmembership{Senior Member,~IEEE}}
\thanks{The authors are with the Centre for IoT and Telecommunications, School of Electrical and Information Engineering, The University of Sydney, NSW 2006, Australia. 
Emails: \{faisal.nadeem, yonghui.li, branka.vucetic, mahyar.shm\}@sydney.edu.au.}}

\maketitle
\begin{abstract}
This paper analyzes wireless network control for remote estimation of linear time-invariant dynamical systems under various Hybrid Automatic Repeat Request (HARQ) packet retransmission schemes. In conventional HARQ, packet reliability increases gradually with additional packets; however, each retransmission maximally increases the Age of Information and causes severe degradation in estimation mean squared error (MSE)  performance. We optimize standard HARQ schemes by allowing partial retransmissions to increase the packet reliability gradually and limit the AoI growth. In incremental redundancy HARQ, we optimize the retransmission time to enable the early arrival of the next status updates. In Chase combining HARQ, since packet length remains fixed,  we allow retransmission and new updates in a single time slot using non-orthogonal signaling. Non-orthogonal retransmissions increase packet reliability without delaying the fresh updates. We formulate bi-objective optimization with the proposed variance of the MSE-based cost function and standard long-term average MSE  cost function to guarantee short-term performance stability. Using the Markov decision process formulation, we find the optimal static and dynamic policies under the proposed HARQ schemes to improve  MSE performance further. The simulation results show that the proposed HARQ-based policies are more robust and achieve significantly better and more stable MSE performance than standard HARQ-based policies.

\end{abstract}
\section{Introduction}
\label{sec:Introduction}
The previous generations of cellular communications have primarily focused on increasing spectral efficiency for enhanced mobile broadband (eMBB) services \cite{varrall20165g}. Whereas, the fifth generation (5G)  mobile communication envisions supporting mission-critical services, e.g., industrial automation, tactile internet, smart grid, telesurgery, virtual reality, etc. \cite{antonakoglou2018toward}. These services require very high reliability and low latency for in-time packet delivery. Most of these emerging mission-critical applications, for example,  industrial automation and telesurgery, require remote estimation of the states of the underlying dynamic process over a wireless link \cite{peng2013event}. Most of the existing work on remote estimation assumes a perfect channel and focuses on designing optimal control. However, the wireless channel can deteriorate the packet reliability leading to instability of the control system. Considering the nature of the application, communication and control should be designed simultaneously \cite{peng2021sensing}.

The  eMBB services heavily rely on average packet error rate (PER) reliability with little focus on latency performance. However, in  mission-critical applications that require remote estimation, the PER reliability is linked with its in-time packet delivery \cite{bennis2018ultrareliable}. This is because the packets contain the state of the remote process to be estimated at the receiver \cite{peng2021sensing}.  Therefore, often in-time packet availability at the cost of less reliability could be more beneficial than a more reliable but outdated packet. The age-of-information (AoI) is  the measure of time elapsed between the moment a measurement is generated from the sensor and the moment it becomes available to the receiver after estimation. The AoI is an important metric to track  the freshness of information  \cite{yates2021age}. 
If the sensor always updates the remote-side controller with equal reliability, minimization of AoI leads to the lowest estimation error. This is usually the case when the dynamic process is highly uncorrelated between consecutive status updates \cite{gupta2010estimation}. 

\textcolor{black}{ Most of the time, the dynamic process to be controlled shows some correlation between its status updates \cite{ornee2021sampling,roth2020remote}. Therefore, the  sensor's raw measurements of the dynamic process are first used to estimate the state of the system with estimators such as  the Kalman filter. Then the states are delivered in the form of packets \cite{antonakoglou2018toward}, which carry states that are correlated. Therefore,  if a fresh status update does not arrive in time, the remote estimator can estimate the new status from previously received status updates.}  Furthermore, in a real-time remote estimation of a correlated dynamic process, a fine balance between reliability and freshness should be maintained \cite{Rajaraman2021NotJA}. For example, when the system is slowly evolving, a more reliable old update can better estimate the next state than a less reliable fresh update. In these situations, many researchers suggest retransmission with performance metrics that are non-linear in terms of AoI, such as the value of information and estimation mean squared error (MSE) \cite{Rajaraman2021NotJA,wang2020value,huang2020real}.

\textcolor{black}{
In \cite{gupta2010estimation}, the author proposed Automatic repeat request (ARQ)-based policies to resend failing updates for better remote estimation. Since with ARQ, the receiver does not take advantage of retransmission in increasing the transmission reliability; therefore the optimal ARQ-based policy is to send a fresh update in each transmission. In the fixed hybrid automatic repeat request (HARQ)-based policy,  when the transmitter receives an acknowledgment (ACK) of packet success, it sends a fresh update; otherwise, it retransmits the old update. Unlike ARQ, in fixed HARQ, the receiver can combine the retransmissions with failed packets to increase the reliability of state updates. However, it is observed that often fewer retransmission rounds with little retransmissions are required when minimizing the average state estimation performance of the dynamic process. The authors in \cite{huang2020real} optimize the fixed-HARQ method to obtain policies based on the knowledge of the dynamic process  termed standard HARQ. HARQ has two common types: Chase combining HARQ (CC-HARQ) and incremental redundancy HARQ (IR-HARQ). In CC-HARQ, the whole packet is repeated, and by utilizing maximum ratio combining (MRC) at the receiver, the reliability is increased. With IR-HARQ, the transmitter increases code redundancy by sending a long codeword in chunks. The receiver improves reliability by improving the decoding performance with a longer codeword after each retransmission \cite{zhang2020energy}. In standard-HARQ in \cite{huang2020real}, reliability improves at the expense of increased latency to fresh updates with each retransmission. This often leads to excessive AoI growth, especially when retransmission is not optimized even at finite blocklength (FBL) regime \cite{nadeem2021analysis,faisalMDPI}. Short packet lengths improve the AoI but cause more packet dropouts, and HARQ retransmission becomes necessary. Yet, standard IR-HARQ  and CC-HARQ provide reliability at the cost of AoI penalty due to retransmission overhead \cite{nadeem2022real}. In \cite{huang2020real}, authors attempt to reduce the AoI penalty by limiting the retransmission rounds. Still, the retransmission process of HARQ is not optimized that can provide optimal retransmission without excessive AoI growth.}

\textcolor{black}{
This paper expands on the standard HARQ-based policies proposed in \cite{huang2020real}, where the sensor has only two choices at any available time slot, i.e., repeat an entire packet of old updates or send fresh updates instead. This limit of choices due to the inherent HARQ process leads to poor long-term average MSE performance over many time slots. Moreover, it results in high MSE performance variation in each time slot due to inappropriate retransmission. All the previous studies on wireless networked control focus only on minimizing the long-term average MSE performance and did not consider the variation of MSE performance \cite{huang2020real}. This means that while the average MSE might be at the desired level, the variation of MSE in each time slot can cause the system to become unstable. In this paper, we highlight the limitation of  standard HARQ in achieving better MSE and propose novel HARQ techniques  tailored to the needs of real-time wireless remote estimation. We consider the variation of MSE in each time slot and try to minimize that as an objective along with long-term average MSE in our proposed optimization using  novel HARQ techniques. 
The main contributions of the paper are summarized as follows:}
\begin{itemize}
\item
We consider the variance of MSE as a cost function and minimize it jointly with the  long-term average MSE cost function. This is to ensure stable MSE performance over short periods, which is ideal for time-critical systems. 
\item
We show the sensitivity of various HARQ schemes in providing stable MSE performance. We highlight that standard HARQ techniques suffer from  MSE performance degradation due to their rudimentary packet  retransmission mechanism that involves delaying new packets. This  could result in  higher MSE variation  due to excessive retransmission.
\item
We propose IR-HARQ and CC-HARQ-based packet retransmission schemes to provide better and more stable MSE performance for a real-time remote estimation system. In IR-HARQ, we optimize the retransmission fraction $\tau$ according to the varying correlation of the underlying dynamic process. Optimal retransmissions enable quicker status updates, limit AoI growth, and improve the MSE performance. In CC-HARQ, since the complete packet is repeated, we propose non-orthogonal CC-HARQ (N-CC-HARQ) to implement incremental age and reliability growth. N-CC-HARQ uses superposition coding with power fraction $\alpha$ to support the simultaneous transmission of fresh updates and retransmission of an old status update in a single time slot. 
\item
We further formulate the policy optimization problem, where the best $\tau$ or $\alpha$ can be chosen with IR-HARQ or N-CC-HARQ, respectively, to minimize the long-term average MSE and its variance jointly. 
The underlying policy optimization problems are complex due to multi-objective criteria and other factors. We use the Markov decision process (MDP) framework to obtain the optimal policy iteratively using numerical techniques. 
\item
Next, we show the impact of $\alpha$ on the system performance, present a static and dynamic setting of $\alpha$, and find the optimal policy. In a static-optimal policy, the optimal  $\alpha$ remains fixed for all time slots. We further enhance the N-CC-HARQ performance by designing a dynamic-optimal policy. With a dynamic-optimal policy, the sensor can select a non-orthogonal power fraction $\alpha$ in each time slot. The dynamic action increases flexibility and leads to better MSE performance than static-optimal policy. We perform intensive simulations to study the impact of different design variables such as process correlation, $\tau$, and $\alpha$, etc. 
\end{itemize}
The rest of the paper is organized as follows. The system model and preliminaries on wireless remote estimation of linear time-invariant (LTI) system with HARQ in the FBL regime and performance metrics are given in Section \ref{Sec:System_Model}. In Section  \ref{Sec:Optimizaiton_HARQ}, we present a detailed analysis of proposed  IR-HARQ and CC-HARQ-based schemes using MDP. We provide numerical results in Section \ref{Sec:Numarical_resutls}. Section \ref{Sec:Practical_Considerations} highlights some issues regarding the practical implementation of the schemes. Finally, Section \ref{sec:conclusion} concludes the paper.
\section{System Model and Preliminaries}
\label{Sec:System_Model}
\begin{figure}[t]
\centering
\includegraphics[width=0.8\columnwidth]{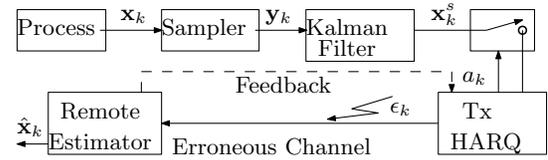}
\vspace{-0.22cm}
\caption{The system model for remote estimation of a dynamic process using HARQ over a wireless channel.}
\label{fig:systemmodel}
\end{figure}

Similar to \cite{huang2020real},  we assume that a smart sensor periodically samples the dynamic process, performs local estimates using the Kalman filtering, and  sends the local estimates over the wireless links. As illustrated in Fig. \ref{fig:systemmodel}, the sensor utilizes retransmission  using HARQ to increase reliability.  
\subsection{Dynamic Process Modeling}
We model the dynamical process with a general discrete LTI system  given as  (e.g., \cite{schenato2008optimal,liu2016energy})
\begin{align}
\mathbf{x}_{k+1}=\mathbf{A}\mathbf{x}_{k}+\mathbf{w}_k, \nonumber \\
\mathbf{y}_k=\mathbf{C}\mathbf{x}_k+\mathbf{v}_k,
\label{eq:LTI}
\end{align}
where the sampling period of the sensor $T_s$ determines the discrete time steps denoted by index $k$. Vector $\mathbf{x}_k\in \mathbb{R}^r$ contains the states of the dynamic process that varies according to state transition matrix $\mathbf{A}\in \mathbb{R}^{r\times r}$. The sampler collects $u$ measurements denoted as   $\mathbf{y}_k\in \mathbb{R}^u$ according to measurement matrix $\mathbf{C}\in \mathbb{R}^{r\times u}$ for local state estimation as shown in Fig. \ref{fig:systemmodel}. The
$\mathbf{w}_k\in \mathbb{R}^r$ and $\mathbf{v}_k\in \mathbb{R}^u$   are
identically distributed (i.i.d.) zero-mean Gaussian processes and measurement noise vectors 
with corresponding covariance matrices $\mathbf{Q}_w$ and $\mathbf{Q}_v$ respectively. 
The initial state $\mathbf{x}_0$ is zero-mean Gaussian with covariance matrix $\mathbf{\Sigma}_0$.
\subsection{Local State Estimation}
The sensor collects raw measurements that are noisy. However, with some storage and computation, a sensor can process the raw data to get state estimation of a process $\mathbf{x}_k$. This can be done using the state-of-the-art Kalman filter \cite{liu2016energy} to get minimum MSE in state estimation from current and previous raw measurements as follows:
\begin{subequations}
\begin{align}
    \mathbf{x}_{k|k-1}^s&=\mathbf{A}\mathbf{x}_{k-1|k-1}^s\\
    \mathbf{P}_{k|k-1}^s&=\mathbf{A}\mathbf{P}_{k-1|k-1}^s\mathbf{A}^{\mathrm{T}}+\mathbf{Q}_w\\
\mathbf{K}_k&= \mathbf{P}_{k|k-1}^s\mathbf{C}^{\mathrm{T}}(\mathbf{C}\mathbf{P}_{k|k-1}^s\mathbf{C}^{\mathrm{T}}+\mathbf{Q}_v)^{-1}\\
\mathbf{x}_{k|k}^s&=\mathbf{x}_{k|k-1}^s+\mathbf{K}_k(\mathbf{y}_k-\mathbf{C}\mathbf{x}_{k|k-1}^s)\\
\mathbf{P}_{k|k}^s&=(\mathbf{I}-\mathbf{K}_k\mathbf{C})\mathbf{P}_{k|k-1}^s,
  \end{align}
\end{subequations}
where at time $k$, $ \mathbf{x}^s_{k|k-1}$ and $ \mathbf{x}^s_{k|k}$ denotes the \textit{priory} and \textit{posteriori} state  estimates, respectively,  whereas, $\mathbf{P}_{k|k-1}^s$ and $\mathbf{P}_{k|k}^s$ are the \textit{priory} and \textit{posteriori} error covariance matrices, respectively. With the Kalman gain, $\mathbf{K}_k$, first two equations are used to perform prediction, whereas rest are used to update the estimates \cite{maybeck1982stochastic}. 
\textcolor{black}{Also, we assume that Kalman filter converges exponentially fast to steady state error covariance. This guarantees the stability of the local estimation asymptotically, i.e., $\lim_{t\to\infty}\mathbf{P}^s_{k|k} = \bar{\mathbf{P}}_0$ \cite{rhudy2013online}, where $\bar{\mathbf{P}}_{0}$ represent the converged error covariance matrix.}
We focus on the performance of remote estimation only with the assumption that parameters $(\mathbf{A},\mathbf{C})$ of the  LTI system in \eqref{eq:LTI} are observable while parameters $(\mathbf{A},\sqrt{\mathbf{Q}}_w)$ are reachable. 

\subsection{Channel Model}
 Let $s(t)$ is the transmitted and $y(t)$ denote the received signal at time $t$ given as:
\begin{align}
y(t)=s(t)+w(t),
\end{align}
where  $w(t)\sim\mathcal{CN}(0,N_0)$ is the circularly symmetric zero-mean complex additive white Gaussian noise (AWGN). We also assume that the total transmit power is $\mathbb{E}[|s(t)|^2]=P_t$ and $s(t)$ is linearly modulated and transmitted with normalized symbol rate 1 symbol/s/Hz. Note that we assume $N_0=1$. Each information packet of $b$ message bits is encoded into a codeword of length $n$ and transmitted using HARQ in $m$ retransmission rounds. We further assume perfect ACK and negative-ACK (NACK) signaling without any delay.
\subsection{HARQ-based Communication over AWGN channel}
The sensor's local estimation is first quantized into $b$ message bits and encoded into a packet carrying $n$ symbols, each with  $\mathrm{T}'_{s}$ duration. 
The sampling period of the sensor is set to be $\mathrm{T}_{s}=n\mathrm{T}'_{s}$ so that the sensor  samples at packet duration. Therefore, a new sample update is collected after delivering one packet of duration $n\mathrm{T}'_{s}$. In each time slot, the sensor can  send a fresh update  repeat the old update.  In HARQ, the sensor sends a coded packet in multiple time slots, and the decoder can combine the repeated transmissions to improve packet reliability.

 HARQ improves packet reliability with each retransmission in two different ways, i.e., incremental signal-to-noise-ratio (SNR) or incremental redundancy known as CC-HARQ and IR-HARQ, respectively. In CC-HARQ, the full coded packet of length $n$ is repeated for retransmission purposes and combined with MRC  followed by a decoding decision. \textcolor{black}{In CC-HARQ, the codeword length $n$ remains fixed, while the effective SNR improves with each retransmission. In IR-HARQ, the transmitter encodes $b$ information bits using $n$ length channel code to be transmitted in maximum $m$ transmission
attempts, e.g.,  $n=\sum_{i=1}^m n_i$ and $n_i$ represents the packet length in the $i$-th transmission attempt. For example, $n_1$ is the packet length with the first transmission with $n_1-b$ bits redundancy. If the  packet reliability is not satisfactory, the IR-HARQ-based transmitter sends additional $n_2$ length redundancy. With each  retransmission, more redundancy is released, until the packet is decoded successfully  or maximum $n$ symbols are sent.} IR-HARQ increases the codeword length with each retransmission, thus increasing the robustness of the error correction performance with longer codewords \cite{erseghe2016coding,nadeem2021non}. 

We assume that the sensor utilizes  short packets to send each update.
 Using normal approximation (NA) \cite{polyanskiy2010channel}, the error rate  of CC-HARQ and IR-HARQ with $m$ cumulative transmissions in the FBL can be respectively characterized as:\footnote{ \textcolor{black}{The NA accurately predicts the performance at short lengths.  Also, in \cite{sahin2019delay}, the author develops practical  HARQ schemes with actual Luby transform (LT) to match the performance at code length $n=100$ and $m=2$.} }

\begin{align}
   {\epsilon}_\mathrm{cc}\left([\gamma_i]_1^{m}\right)\approx Q\left( \frac{n \log_2(1+\sum_{i=1}^{m} \gamma_i)-b+\log_2(n)}{n \sqrt{   V(\sum_{i=1}^{m} \gamma_i)}}\right),
    \label{eq:CC-HARQ_AW}
 \end{align}
\begin{align}
\nonumber{\epsilon}_\mathrm{ir}&\left([\gamma_i]_1^{m}, [n_i]_1^{m}\right)\approx \\
    &Q\left( \frac{\sum_{i=1}^{m} n_i \log_2(1+\gamma_i)-b+\log_2(\sum_{i=1}^{m} n_i)}{\sqrt{\sum_{i=1}^{m}  n_i V(\gamma_i)}}\right),
    \label{eq:IR-HARQ_AW}
 \end{align}
\textcolor{black}{where $n_j=\tau n_1$ for $j=[2, \cdots m]$ with $\tau=(0-1]$,}  $[x]_1^m=[x_1,\cdots,x_m]$, $V(\gamma_i)= \left(1-(1+\gamma_i)^{-2}\right) \log_2^2(e)$ is the channel dispersion, $\gamma_i$ is the SNR at the $i$-th transmission round, and $Q(.)$ is the standard $Q$-function. For IR-HARQ,  $0\leq\tau\leq 1$ is used to select the redundancy $n_j$ with respect to $n_1$.
 
\subsection{Remote State Estimation}
We assume that the sensor continuously updates the receiver by sending packets in each time slot. If a packet fails in the current time slot, the receiver estimates the state based on the old successful packets in the most immediate previous time slots. However, a current state estimate based on the old update leads to poor estimation due to higher AoI.  We assume that one unit packet transmission delay is equivalent to a unit time slot that exists between receiver and the sensor. Thus a sensor's measurement at time slot $k$ is available to the receiver before time slot $k+1$. If the packet fails, it delays the estimation by unit time slot, i.e.,  AoI increases by one. More specifically, if the sensor's local estimate generated at the time instant $t_k$, denoted as $\hat{\mathbf{x}}_{t_k}^{s}$, becomes available to the  receiver at the beginning of time slot $k$, the $q_k$ is given as \cite{kaul2012real}
\begin{align}
    q_k = k-t_k, \quad \forall k 
    \label{eq:age_of_information}
\end{align}
where $q_k\geq1$ is the AoI. The receiver estimation quality depends upon the AoI. With AoI measure $q_k$, the optimal MSE  estimator at the receiver at the beginning of time slot $k$  is  \cite{schenato2008optimal}
\begin{align}
\hat{\mathbf{x}}_k=\mathbf{A}^{q_k} \hat{\mathbf{x}}_{t_k}^s,
\label{eq:Opt_estimater}
\end{align}
and the error covariance matrix of the estimation can be defined as \cite{schenato2008optimal}
\begin{align}
\mathbf{P}_k=\mathbb{E}[({\mathbf{x}}_k-\hat{\mathbf{x}}_k)({\mathbf{x}}_k-\hat{\mathbf{x}}_k)^\mathrm{T}],
\label{eq:covarience_k}
\end{align}
where putting \eqref{eq:LTI} and \eqref{eq:Opt_estimater} into \eqref{eq:covarience_k}, the  one-to-one correspondence of remote estimation error covariance with the AoI $q_k$ is obtained as
\begin{align}
\mathbf{P}_k=f^{q_{k}}(\bar{\mathbf{P}}_{0}),
\label{eq:covarience_aoi}
\end{align}
\textcolor{black}{where  $f(\mathbf{X})\triangleq \mathbf{AXA}^{T}+\mathbf{Q}_{w}
$ and $f^{q+1}(\cdot) \triangleq f (f^{q}(\cdot))$, $f^{1}(\cdot) \triangleq f(\cdot)$. Then, the instantaneous cost associated with the remote estimation error at AoI $q_k$ is $\mathrm{Tr(\mathbf{P}_k)}$. According to \cite[Lemma 3.1]{shi2012scheduling} $\mathrm{Tr(\mathbf{P}_k)}$ monotonically increases with $q_k$, i.e.,
$\mathrm{Tr}\left ({f^{q}(\bar{\mathbf{P}}_{0})}\right) \leq \mathrm {Tr}\left ({f^{q+1}(\bar{\mathbf{P}}_{0})}\right)$ \cite{wu2020optimal}. The quantity $\rho^2{(\mathbf{A})}$ indicates the correlation of the LTI dynamic process. When $\rho^2{(\mathbf{A})}$ is high, the prediction error relative to AoI increment is higher and vice versa. This is because when the process state changes faster, its estimation with delay is poorer than when the state changes gradually.}

\subsection{Performance Metrics}
\textcolor{black}{The bounded long-term average MSE of the dynamic process $\bar{\mu}_\mathrm{MSE}$ indicates that the remote estimation is mean-square stable. The long-term average MSE cost for wireless remote estimation is defined as: \cite{huang2020real}.
\begin{align}
    \bar{\mu}_\mathrm{MSE}=\limsup _{K\to \infty }\frac {1}{K}\sum _{k=1}^{K} \mathbb{E}_w\left [{\text {Tr}\left ({\mathbf {P}_{k}}\right)}\right],
    \label{eq:metric_average_MSE}
\end{align}
where the expectation  $\mathbb{E}_w$ is taken due to the randomness of the packet failure in each time slot using Monte Carlo simulations.
This cost function is used intensively in the study of status update systems and systems involving wireless networked control (WNC) to identify the average AoI-penalty guarantee \cite{ashok2017value}. Note that with   \eqref{eq:metric_average_MSE}, the accumulated cost over many packets becomes representative as $K$ increases, and the MSE cost variation that might appear infrequently cannot be observed. Therefore, the average performance criteria can diminish  MSE  variations occurring over a shorter duration, especially the one that arise due to excessive retransmissions of standard HARQ methods. Moreover, since our proposed HARQ methods reduce the excessive MSE penalty of standard HARQ in each retransmission, we introduce another performance metric to highlight average MSE variation over time as follows:
\begin{align} 
 \bar{\sigma}^2_{\mathrm{MSE}}= \limsup _{K\to \infty }\frac{\sum_{k=1}^K \mathbb \vert \mathbb{E}_w [\text {Tr} ({\mathbf {P}_{k}})]-\bar{\mu}_\mathrm{MSE} \vert ^2}{K}.
\label{eq:metric_deviation}
\end{align}
 Note that this variation of the performance in each time slot occurs due to packet failure because of short packet lengths and AWGN channel noise according to  \eqref{eq:CC-HARQ_AW} and \eqref{eq:IR-HARQ_AW}. This makes instantaneous MSE $\mathrm{Tr} ({\mathbf {P}_{k}})$ behave randomly with applied HARQ. The quantity $\bar{\sigma}^2_{\mathrm{MSE}}$ compares on average how much the instantaneous MSE in each time slot varies from the mean. }


\section{HARQ-Optimized Transmission control: Design, Analysis and Problem Formulation  }
\label{Sec:Optimizaiton_HARQ}
\textcolor{black}{When the packet reliability remains fixed with or without retransmission, such as in standard ARQ, the optimal policy is to avoid retransmissions so that every time the sensor’s current updates are transmitted  \cite{gupta2010estimation}. Whereas, under HARQ, the retransmissions are more reliable than new transmissions, because the packet reliability improves with each retransmission. Therefore, there exists an inherent trade-off between  sending the current state with low reliability and retransmitting failed old status update with higher reliability. The authors in \cite{huang2020real} discuss this trade-off and introduce an HARQ-based policy where the number of retransmissions is optimized to show improvement in estimation MSE in comparison to ARQ. Using standard IR-HARQ, the authors in  \cite{huang2020real} obtain a policy to send fresh updates or retransmission when a packet fails.  }

\textcolor{black}{In standard  HARQ retransmissions, higher packet reliability is achieved at the cost of increasing the AoI for the duration of retransmission. Therefore, it is important to make sure that HARQ retransmissions provide enough reliability to recover a failing packet and at the same time not in excess to cause additional AoI penalty. Excessive retransmisison overhead will result in waste of resources that could otherwise be used to reduce MSE further. On the other hand,  inadequate retransmission will not stop the packet failure raising the MSE again. Therefore, it is important to optimize the HARQ retransmissions to provide MSE improvement for each status update. Clearly, the optimization of  HARQ retransmission action  will reduce the excessive growth of estimation MSE impacting both cost functions defined in \eqref{eq:metric_average_MSE} and \eqref{eq:metric_deviation}. The design of sensor's action policy with optimized IR-HARQ and CC-HARQ is given next.}


%






\subsection{Proposed IR-HARQ Design for WNC}
Let $a_k=\{0,1\}$ denote the sensor's control action of sending a fresh update or retransmitting an old update in time slot $k$ respectively. \textcolor{black}{In standard IR-HARQ method adopted in \cite{huang2020real}, when packet of length $n_1$ is successful with action $a_k=0$ the AoI is 1 unit. This is because the normalized AoI for length $n_1$ packet is considered $1$.  When the sensor takes action $a_k=1$, it sends additional redundancy of length $n_2=n_1$ increasing the AoI further  by $1$ unit \cite{huang2020real}. In most practical cases, often packet fails due to a few corrupted bits and requires less redundancy to recover. We utilize the inherent  capability of the IR-HARQ method to adopt the amount of redundancy  to reduce the time  for each retransmission. The shorter retransmission slot will enable decoding of old status update with the least AoI as well as make sure that the next status update is delivered  earlier.}  In the proposed IR-HARQ method, we  introduce fraction $\tau$ to control the amount of redundancy during retransmission, e.g., $n_2=\tau n_1$.

\subsubsection{Transmission control policy with \texorpdfstring{$0\leq \tau \leq 1$}{Lg} }
\begin{figure}[t]
\centering
\includegraphics[width=0.8\columnwidth]{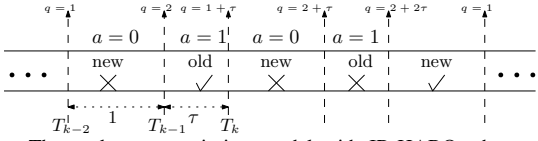}
\vspace{-0.15in}
\caption{The packets transmission model with IR-HARQ when maximum single retransmission with $\tau$ fraction is allowed.}
\label{fig:pcketization-IR}
\end{figure}

    \textcolor{black}{As show in Fig. \ref{fig:pcketization-IR}, the action taken at each time slot  is denoted as $a\in\{0,1\}$. The $k$-th time slot duration is given as $T_{k-1} - T_{k}$, which also indicates the sampling interval. During the time slot $k-1$, the sensor updates the controller with the fresh update with the action $a=0$. If the receiver is unable to decode the packet, the sensor sends the retransmission according to fraction $\tau$ related to the old update in time slot $k$ with action $a=1$. }
\textcolor{black}{The parameter $m_k$ denotes the number of transmissions  for delivering single status before time slot $k$. For example, after the first retransmission, $m=2$. $m_k$ depends on the transmission control policy with HARQ given as:
\begin{align}
m_{k} = \begin{cases} 
 1, & \mathrm{when} \quad a_{k-1}=0,\\ 
m_{k-1}+1 , & \mathrm{when} \quad a_{k-1}=1.
\end{cases}
\label{Eq:m_updating}
\end{align}
For IR-HARQ, packet reliability ${\epsilon}_\mathrm{ir} \left([\gamma_i]_1^{m}, [n_i]_1^{m}\right)$, varies according to \eqref{eq:IR-HARQ_AW} due to changing $n_i$ in proportion to selected $\tau$. For example at $m=2$,  $n_2=\tau n_1$.}
 %
  
Similarly, the AoI  ($q_k$)  with $m$ partial retransmissions under $\tau$, is given as 
\begin{align}
q_{k} = \begin{cases} 
1, & a_{k-1} = 0, \zeta _{k-1}  =0,\\
q_{k-1}+1, & a_{k-1} = 0, \zeta _{k-1}  =1,\\ 
(m_{k-1}-1)\tau+1, & a_{k-1} = 1, \zeta _{k-1} = 0,\\ 
q_{k-1}+(m_{k-1}-1)\tau+1, & a_{k-1} = 1, \zeta _{k-1}  =1, 
\end{cases}
\label{Eq:age_updating}
\end{align}
where  $\zeta _{k} =0$ and $\zeta _{k} =1$ indicates the packet successful and fail  states at the receiver during time slot $k$, respectively. The packet reliability for IR-HARQ at various $m$ with specific $\tau$ is varied as \eqref{eq:IR-HARQ_AW}. Therefore  equations \eqref{eq:IR-HARQ_AW} \eqref{Eq:m_updating} and \eqref{Eq:age_updating} links the sensor's action with its AoI penalty under finite block length IR-HARQ based communication with $\tau$. 

\textcolor{black}{Finally, the instantaneous AoI $q_k$ can be used in  \eqref{eq:covarience_aoi} to obtain an exact MSE cost with IR-HARQ at specific $\tau$ point. The AoI scales linearly with time according to \eqref{eq:age_of_information}, whereas estimation MSE is the non-linear penalty function of AoI. Furthermore, the relationship between $\mathrm{Tr(\mathbf{P}_k)}$ and $q_k$ for specific system dynamics involving state matrix $\mathbf{A}$ is derived in \cite[Lemma 3.1]{shi2012scheduling}. Using this relationship, we can characterize the MSE cost at discrete values of  AoI. In \cite{shi2012scheduling, huang2020real}, the AoI takes integer values due to fixed packet length. However,  the MSE  varies, according to correlation in the dynamic process leading to different MSE cost penalties at fixed AoI. This is achieved by scaling the AoI to the appropriate MSE with respect to $\mathbf{A}$. We also use this scaling to map the fractional growth of AoI to its appropriate MSE penalty at specific $\mathbf{A}$. More specifically, the estimation MSE cost grows exponentially with a rate proportional to quantity $\rho^2(\mathbf{A})$, which is the maximum squared eigenvalue of $\mathbf{A}$. The matrix $\mathbf{A}$ of LTI in \eqref{eq:LTI} are properties of the system and are determined by the system structure and elements. For different values of $\rho^2(\mathbf{A})$ we also consider different penalty functions, which makes this scaling accurate.}

\subsubsection{Problem Formulation}
\textcolor{black}{For a given $\tau$, the sensor can send a fresh update or retransmission of an old update in each time slot $k$. The optimal $\tau$ limits the excessive MSE cost penalty due to retransmission leading to less variation of MSE in each time slot. Let $\lambda(\tau)\triangleq\{ a_1, a_2, \cdots, a_k, \cdots \}$  denotes the sensor action policy indicating the sequence of actions, where $a_k=\{0,1\}$ changes the reliability and AoI with respect to $\tau$  according to \eqref{Eq:m_updating} and \eqref{Eq:age_updating}.    
 The goal is to find the optimal policy $\lambda(\tau)$  so that the long-term estimation MSE and its variance are minimized over $K$ time slots. The problem formulation is as follows:  
 \begin{equation}
 \begin{aligned}  
\textrm{Prob. 1}:  \underset{\lambda(\tau) }{\textrm{minimize}}  & \; {(\bar{\mu}_\mathrm{MSE},                     \bar{\sigma}^2_{\mathrm{MSE}}  )}\\
\textrm{s.t.} \quad\textrm{C}_1: & \;  0\leq\tau\leq 1
\label{eq:optimal_policy_IR_HARQ}
\end{aligned}
 \end{equation}
where,  $\bar{\mu}_\mathrm{MSE}$ and $\bar{\sigma}^2_{\mathrm{MSE}}$ are long-term average MSE and its variance defined by \eqref{eq:metric_average_MSE} and \eqref{eq:metric_deviation}, respectively.
The problem \textrm{Prob.1} is a complex bi-objective optimization problem (BOOP) whose closed-form solution is not tractable for global optimization \cite{branke2008multiobjective}. We find the Pareto optimal solution using the $\epsilon$-constraint method first proposed by Haimes et al. in 1971 \cite{haimes1971bicriterion}. 
 In this method, one of the objective functions is selected to be optimized while the other(s) are converted into constraints, leading to a solution that is proven to be always weakly Pareto optimal \cite{branke2008multiobjective}. This method is intuitive and flexible to allow some modification based on the nature of the objective functions and priorities of one over the other. For example, we keep  $\bar{\mu}_\mathrm{MSE}$ as the main objective of the transformed problem while treating \eqref{eq:metric_average_MSE} as a constraint. One reason for this choice is that such problems can be solved using the MDP framework. The $\epsilon$-constraint problem of the Prob.1 is given as follows:
 \begin{equation}
    \begin{aligned}  
\textrm{Prob.1a}: \underset{\lambda(\tau) }{\textrm{minimize}}  & \; {\bar{\mu}_\mathrm{MSE}}\\ 
 \textrm{s.t.} \quad  \textrm{C}_1:& \; 0\leq\tau\leq 1 \\ 
              \quad  \textrm{C}_2:&  \;  \bar{\sigma}^2_{\mathrm{MSE}}(\lambda(\tau)) \leq \theta 
\label{eq:IR_HARQ_transformation}
\end{aligned} 
 \end{equation}
where the $\theta$ denotes the value of $\epsilon$-threshold} \footnote{We use  $\theta$ instead of standard $\epsilon$ symbol to avoid confusion between threshold and the error symbol used in \eqref{eq:CC-HARQ_AW} and  \eqref{eq:IR-HARQ_AW}}. 
\subsubsection{Numerical solution}
\textcolor{black}{We use a heuristic approach to solve the $L$ different versions of this problem, each with  a specific level $\theta_{\ell}$  \cite{chiandussi2012comparison}. We obtain various threshold levels using the dependency of $\bar{\sigma}^2_{\mathrm{MSE}}$ on the system parameter $\tau$. For that we use $L$ discrete values of $\tau$ as $\{\tau_1, \tau_2, \cdots \tau_{L}\}$. Then we can set the value of $\theta_\ell$ using worst-case policy, i.e., sending maximum retransmission for each status update $\lambda(\tau_{\ell})$ for $K$ slots using \eqref{eq:metric_deviation}. This gives the upper limit of $\theta$ for given $\tau$. The next step is to find the optimal policy $\lambda^{*}(\tau_{\ell})$ that minimizes the long-term average MSE for the selected value of $\tau_{\ell}$ under the constraint $\textrm{C}_2$. Then for each  $\tau_{\ell}$, this problem becomes a classic average cost optimization problem under additional constraint $\textrm{C}_2$. For a fixed value of $\tau$, \eqref{eq:covarience_k}, \eqref{Eq:m_updating} and \eqref{Eq:age_updating} show that the states $q_k$ and $m_k$ and cost $\mathrm{Tr}{(\mathbf{P}_k)}$ depends only on  states and actions in previous time slots, i.e., $m_{k-1}$, $q_{k-1}$ and action $a_{k-1}$. This Markov property allows to transformation of the problem $\textrm{Prob.1a}$ into discrete time MDP for specific $\tau$ value.}

\subsubsection{MDP Formulation} 
\label{Sec:MDP_IR}
  In general, an  MDP is defined by state space, action space, state transition function and reward function (cost function) denoted as $\mathbb {J}$, $\mathbb {A}$, $\mathbb {P}({J}_{k}\vert {J}_{k-1}, a)$ and $\mathbb {C}({J}_{k}, a)$, respectively. The components are defined as: \textcolor{black}{ \textbf {(a) The state space} which is $\mathbb{J}\triangleq\{ (m,q),\in \mathbb{N}\times \mathbb{R}^{+}\}$, such that   ${J}_k\triangleq(m_k,q_k)\in\mathbb {J}$ denotes the state of the MDP at time $k$. The $\mathbb{N}$ and $\mathbb{R}^{+}$ indicates the natural and real positive number space according to  \eqref{Eq:m_updating} and \eqref{Eq:age_updating} respectively for $m$ and $q$.  }\textbf{(b) The action space} of the MDP is $\mathbb {A}\triangleq\{0, 1\}$, where  action $a=0$ corresponds to the sensor sending a fresh update using length $n_1$ packet. The action $a=1$ corresponds to the sensor   retransmitting an old update with packet length $n_2=\tau n_1$, where $0\leq\tau\leq 1$. Selecting higher $\tau$ increases packet reliability  at the cost of a rapid increase in AoI  with each retransmission. \textbf{(c) The state transition function} $\mathbb {P} ({J}_{k+1}\vert {J_{k}}, a)$ is the probability of state transition from  two consecutive states. As the transition is time-homogeneous, we can drop the time index $k$ and denote the current state and next state as ${J}$ and ${J}^{'}$, respectively. Consequently, the number of retransmission  $m$ and AoI $q$ increased  according to \eqref{Eq:m_updating} and \eqref{Eq:age_updating}, respectively. For given $m$,  $q$ and $\tau$, the state transition function is given as 
\begin{align} 
\mathbb {P} ({J}^{'}\vert {J}, a)  =
\begin{cases} 
1-\epsilon_\mathrm{ir}^{(m)}, & {J}^{'}= \big(m,(m-1)\tau+1\big), \\ 
\epsilon_\mathrm{ir}^{(m)}, & {J}^{'}= \big(m,q+(m-1)\tau+1\big), \\ 
0, & \text {otherwise},\\ 
\end{cases} 
\label{eq:state_transition_IR}
\end{align}
where $\epsilon_\mathrm{ir}^{(m)}$ is the  concise notation corresponding to  the error probability  of IR-HARQ in finite block length at the $m$-th transmission given in \eqref{eq:IR-HARQ_AW} as $\epsilon_\mathrm{ir}([\gamma_i]_1^m,[n_i]_1^m)$.  \textbf{(d) The cost function} associated with each action is the instantaneous MSE at the current state. The cost function is a non-linear function of AoI  $q$ given as 
\begin{align}
    \mathbb{ C}({J},a) \triangleq \text {Tr}\left ({f^{q}(\bar {\mathbf {P}}_{0})}\right).
    \label{eq:cost_function_IR}
\end{align} 
The above MDP is solved by selecting discrete $\tau_\ell$ value leading to a specific policy $\lambda(\tau_{\ell})$.  
\textcolor{black}{It can be seen that $m$ and $q$ can increase unbounded as \eqref{Eq:m_updating} and \eqref{Eq:age_updating} resulting in infinite state space $\mathbb{J}$. Therefore, we truncate the state space for numerical implementation with a certain $q_\mathrm{max}$ value. As a result, in the event of error at state $J(m,q_\mathrm{max})$ state transits to itself with probability $\epsilon_\mathrm{ir}^{(m)}$ in state transition equation \eqref{eq:state_transition_IR}.} Also, we assume that the value of $\tau_\ell$ does not change across time slots. Due to the exponential growth of the cost function with $q$, it is possible that the HARQ-based policy $\lambda(\tau_{\ell})$ cannot be bounded. This can happen if the packet error rate in relation to the state change is significantly higher. However, a simple sufficient condition $\epsilon_\mathrm{ir}^{m}\rho^2(\mathbf{A})< 1$ guarantees that the optimal policy exists that achieve bounded long-term average MSE cost as proved in \cite[Theorem 1]{huang2020real}. \textcolor{black}{Also, notice that each state of the MDP can be visited from any other state using the connections through \eqref{eq:state_transition_IR}, where $1-\epsilon_\mathrm{ir}^{(m)}$ indicates the probability of single self transiting loop. Therefore, the  MDP belongs to an a-periodic uni-chain which can be solved using standard value iteration algorithm \cite{littman2013complexity} to obtain $\lambda(\tau_\ell)$}. We solve $L$ such MDPs, each with specific $\tau_{\ell}$ value leading to $L$ policies with corresponding cost values of each objective function. Then the optimal policy can be selected that minimizes both objective out of $L$ policies \cite{fan2016improved}.   

\begin{remark}
The equivalent MDP problem of \cite[Eq.(25)]{huang2020real} can be obtained by fixing maximum retransmissions to 1, i.e.,  $\tau=1$ in \eqref{eq:state_transition_IR}. Therefore, the  IR-HARQ-based approach adopted in \cite{huang2020real} is a special case of the above proposed design in Prob.1a when only a single level of $\theta$ is selected at  $\tau=\tau_1=1$. We refer to it as the standard IR-HARQ method. 
\end{remark}


\subsection{Proposed Non-orthogonal CC-HARQ Design for WNC}
\label{sec:NOMA_MODEL}
\begin{figure}[t]
\centering
\includegraphics[width=0.8\columnwidth]{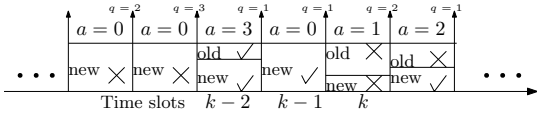}
\vspace{-0.19cm}
\caption{Packet structure with DN-CC-HARQ when $L=3$ and $m=2$}
\label{fig:pcketization}
\end{figure}

In standard CC HARQ, when a packet carrying status update fails, it is repeated in the next time slot to increase the  SNR. This increases the reliability of the status update but at the cost of missing a newly generated status update during retransmission time slots. Furthermore, since in standard CC-HARQ, the complete packet is transmitted occupying a full time slot, it maximally increases the AoI leading to poor MSE performance. To counter this, we propose a novel non-orthogonal CC-HARQ (N-CC-HARQ)  scheme that increases the reliability of an update without stopping the continuous arrival of the fresh status update, as shown in Fig. \ref{fig:pcketization}. With N-CC-HARQ, an old update is retransmitted non-orthogonally with freshly generated status, using appropriate power-sharing fraction $\alpha$. More specifically, $\alpha_k P$ power is assigned to the old update and $(1-\alpha_k)P$ power to the new update during time slot $k$ using superposition coding. After receiving the retransmission, the receiver performs MRC with old copies to increase SNR and separate the overlapping fresh packets using successive interference cancellation (SIC).
As seen in Fig. \ref{fig:pcketization}, that $\alpha_k$ can vary in each time slot; therefore, we call it  \textit{dynamic} non-orthogonal CC-HARQ (DN-CC-HARQ). The DN-CC-HARQ provides  additional flexibility for combining fresh and old updates over many time slots for more controlled AoI growth.


\subsubsection{Transmission Control Policy of DN-CC-HARQ}
Let $a_k\in\{0,\ell\}$ respectively denote the  action of sending a fresh update or retransmitting an old update with power sharing fraction $\alpha_{\ell}, \ell=\{1,2,\cdots, L\}$, where $L$ is the total number of  power-sharing fractions available in each time slot. The number of consecutive retransmission $m$ changes according to \eqref{Eq:m_updating}, where action $a=\ell$ represents retransmission action. 
If a packet scheduled in the $(k-1)$-th time slot succeeds under the action $a_{k-1}=0$, the AoI ($q_k$) is 1. When  $a_{k-1}=\ell$,  the AoI  depends on the  packet success and fail state of both the new and old updates. The  AoI is given as 
\begin{align}
q_{k}(\ell) = 
\begin{cases} 
            1, & a_{k-1} = 0, \zeta _{k-1}  =0,\\
            1, & a_{k-1} = \ell, \zeta^\mathrm{o}_{k-1}(\ell)=0 \;\&\;\zeta^\mathrm{n}_{k-1}(\ell) = 0,\\  2, & a_{k-1} = \ell, \zeta^\mathrm{o}_{k-1}(\ell)=0\;\&\;\zeta^\mathrm{n}_{k-1}(\ell) = 1, \\ 
           q_{k-1}(\ell)+1, & \quad \zeta _{k-1}=1 \;\text{or}\; \zeta^\mathrm{o}_{k-1}(\ell)=1,
\end{cases}
\label{eq:age_progression}
\end{align}
where $\zeta_{k-1}=\{0,1\}$ indicate  success or fail state respectively with action $a_{k-1}=0$.  $\zeta_{k-1}^{\mathrm{o}}(\ell)$ and $\zeta_{k-1}^{\mathrm{n}}(\ell)$  indicate fail, or success states  corresponding to  old  and new updates respectively in a non-orthogonal packet.  We assume the transmitter is informed about success or failed packet decoding  using an error-free and zero-delay feedback signal.  
\subsubsection{Problem Formulation}
\textcolor{black}{Let ${\lambda}_{\mathrm{DN}}$ denote the DN-CC-HARQ-based transmission control policy defined as a sequence of actions taken in each time slot as  $\{a_1, a_2,\cdots a_k,\cdots\}$, where $a_k\in\{0,\ell\}$ is the action in time slot $k$. The appropriate value of $\alpha_{\ell}$  will reduce the impact of retransmission overhead, leading to improved MSE performance. The objective is to find the optimal policy, with the flexibility of $\alpha$ that minimizes the cost functions  $\bar{\mu}_\mathrm{MSE}$ and  $\bar{\sigma}^2_\mathrm{MSE}$,   defined by \eqref{eq:metric_average_MSE} and \eqref{eq:metric_deviation} respectively. 
For a specific SNR, message bits $b$, block length $n$, maximum allowed retransmission $m$ and number of power levels $L$,  optimization problem  is given as
\begin{equation}
\begin{aligned}  
\textrm{Prob.2}: \min_{ \lambda_\mathrm{DN} (\alpha) }  & \; {(\bar{\mu}_\mathrm{MSE}, \bar{\sigma}^2_{\mathrm{MSE}})}\\
\textrm{s.t.} \quad  \textrm{C}_1:&  \; 0\leq\alpha_{\ell}\leq 1,
\label{eq:optimal_policy_CC_HARQ}
\end{aligned}    
\end{equation}
where  $\ell=[1,\cdots, L]$. The Prob.2 has a complex bi-objective form, where each cost function is a complicated function of MSE given in \eqref{eq:metric_average_MSE} and \eqref{eq:metric_deviation}. Furthermore, the one-step cost $\mathrm{Tr(\mathbf{P}_k)}$ is a non-linear function of $q_k$ \eqref{eq:covarience_aoi}, which varies according to $\alpha_{\ell}$ and retransmission count $m_k$. Therefore, the closed form or optimal global solution is not realizable due to the complex optimization problem. We use a heuristic approach to solve such problems as indicated in detail for the problem $\textrm{Prob.1}$ in \eqref{eq:IR_HARQ_transformation} and  transform the problem into a single objective as}
\begin{equation}
    \begin{aligned}  
\textrm{Prob.2a}: \min_{ \lambda_\mathrm{DN}(\alpha_{\ell}) }& \; {(\bar{\mu}_\mathrm{MSE} })\\ 
\textrm{s.t.} \quad  \textrm{C}_1:& \; 0.5\leq\alpha_\ell\leq 1,  \\ 
             \quad   \textrm{C}_2: & \; \bar{\sigma}^2_{\mathrm{MSE}}(\lambda_\mathrm{DN}(\alpha_\ell)) \leq \theta. 
\label{eq:optimal_policy_CC_HARQ_transformed}
\end{aligned}
\end{equation}
\subsubsection{Numerical solution}
\textcolor{black}{ We obtain various levels of threshold $\theta$ using the dependency $\bar{\sigma}^2_{\mathrm{MSE}}$ on the system parameter $\alpha_\ell$. For that we use $L$ discrete values of $\{\alpha_1, \alpha_2, \cdots \alpha_{L}\}$. For example, the upper bound of the threshold level $\theta$ for $\textrm{Prob.2a}$ can be obtained by selecting the worst policy with power fraction $\alpha_{\ell}$ that maximized the MSE in each time slot. In our problem,  this policy is to send maximum retransmission for each status update with full power $\alpha_{1}=1$. The next step is to find the optimal policy $\lambda^{*}(\alpha_{\ell})$ that minimizes the long-term average MSE for smaller values $\alpha_{\ell}$ satisfying constraint $\textrm{C}_2$ by varying levels of $\alpha_{\ell}$ with $L$ degrees of freedom giving rise to various $\theta$-levels. Then for each setting, this problem becomes a classic average cost optimization problem due to a single objective function related to average performance, which can be solved using MDP. We solve the corresponding average cost optimization problems using MDP-based policy optimization for a given $\theta$ as follows: }
\subsubsection{MDP-based solution}
We transform the optimization problem  \eqref{eq:optimal_policy_CC_HARQ_transformed} for each $\theta$-level into an MDP using state, action, transition and  cost functions. The complete state space is defined as $\mathbb{J}\triangleq\{ (m,q(\ell))\in \mathbb{N}\times \mathbb{N}\}$, where ${J}_k\triangleq(m_k,q_k(\ell))\in\mathbb{J}$ is the state at $k$-th time slot. 
%
The action  space $\mathbb{A}\triangleq \{0,\ell\} $. 
The state transition function between current state $J$ at time $k$ to next state $J'$ at time $k+1$ is denoted as  $\mathbb{P}({J'}\vert {J}, a)$  is defined in Lemma 1. 
\begin{lemma} The state transition function $\mathbb{P}(J'\vert J, a)$  when $a_k=0$ is given by
\[\mathbb{P}(J'\vert J,a) = \nonumber\\
\left\{\begin{array}{ll}
1-\epsilon(\gamma_1);  \quad\quad\quad\quad \quad & {J'}=(1,1),\\
\epsilon(\gamma_1);\quad\quad\quad\quad  &{J'}=(1,q+1),\\
\end{array}\right.\]
and when  $a_{k}={\ell_1}$ and $a_{k+1}={\ell_2}$ in current and  next time slot
\[= \nonumber\\
\left\{\begin{array}{ll}
\left(1-\epsilon\left(\gamma_i(\ell_1),\gamma_I(\ell_2)\right)\right)\left(1-\epsilon\left(\gamma_2(\ell_2)\right)\right);&J'=(1,1),\\
\left(1-\epsilon(\gamma_i(\ell_1),\gamma_I(\ell_2))\right)\epsilon(\gamma_2(\ell_2));&J'=(2,2),\\
\epsilon(\gamma_i(\ell_1),\gamma_I(\ell_2));& J'=(2,q+1),\\
0;& \text{else},
\end{array}\right.\]
\textcolor{black}{where $\gamma_i(\ell_1)$ is the SNR in the previous time slot, $i\in\{1,2,e\}$ }  $\gamma_1(\ell_1)=P$,
$\gamma_2(\ell_1)=(1-\alpha_{\ell_1} )P$, and $\gamma_e(\ell_1)=((1-\alpha_{\ell_1} )/\alpha_{\ell_1})P$  $\gamma_2(\ell_2)=(1-\alpha_{\ell_2} )P$,
$\gamma_I(\ell_2)=\frac{\alpha_{\ell_2} P}{1+ (1-\alpha_{\ell_2})P}$, $\ell_1=[0,1,\cdots,L]$ and $\ell_2=[1,\cdots,L].$
\end{lemma}
\begin{IEEEproof}
The state transition probabilities regarding action $a=0$ are obtained directly as packet success and failure with single transmission using \eqref{eq:CC-HARQ_AW}. 
Let the sensor takes action $a=\ell_1$ and $a=\ell_2$ in previous and current time slot using $\alpha_{\ell_1}$ and $\alpha_{\ell_2}$ power fractions respectively\footnote{with slight abuse of notation $\alpha_{\ell_1}$ for $\ell_1=0$ here indicates action $a=0$, by defining $\alpha_0=0$. Also, we omit index $k$ for simplicity as any two states can be linked using Lemma 1}. Then, if the current status update fails to be decoded with a single packet, the receiver can rely on the more reliable old update in the current time slot leading to state $(2, 2)$. This is because, with additional retransmission, the AoI increases as well. When overlapping fresh status update in the current time slot fails, the success probability with single retransmission is
 $\left(1-\epsilon(\gamma_i(\ell_1),\gamma_I(\ell_2))\right)\epsilon(\gamma_2(\ell_2))$. However, if the receiver recovers the overlapping fresh status update after removing old update using SIC. the state $(1, 1)$ is reached. The probability of such and event when both signal are recovered is $\left(1-\epsilon\left(\gamma_i(\ell_1),\gamma_I(\ell_2)\right)\right)\left(1-\epsilon\left(\gamma_2(\ell_2)\right)\right)$. Note that we assume $\alpha_i\geq 0.5 \forall i$, for simplification and imposing fixed SIC decoding order. Thus, when a status update fails to be decoding under action $a_k=\alpha_{\ell}$ after a single retransmission, the overlapping fresh status update with single retransmission can never be recovered. The event's probability is $\epsilon(\gamma_i(\ell_1),\gamma_I(\ell_2))$ with retransmission count as $m=2$ and AoI increases as $q+1$.

\textcolor{black}{The $\gamma_i(\ell_1)$ indicate the S(I)NR of the old status update before its retransmission arrive. Due to non-orthogonal transmission, $\gamma_i(\ell_1)$ can take three different value for $i=\{1,2,e\}$. $\gamma_1(\ell_1)=P$, when there was no overlapping transmission occur, $\gamma_2(\ell_1)=(1-\alpha_{\ell_1})P$ when overlapping packet is removed due to SIC and $\gamma_e(\ell_1)=((1-\alpha_{\ell_1})/\alpha_{\ell_1})P$ when SIC fails.  When the sensor takes action $a=0$, the fresh update is available to the receiver with SNR $P$.} Under action $a=\ell_2$ with DN-CC-HARQ, the signal to interference and noise ratio (SINR) of the overlapping fresh update and the retransmitting old update vary with specific $\alpha_{\ell_2}$. For example  the SINR for the retransmission of the old update is $\gamma_I(\ell_2)=\frac{\alpha_{\ell_2} P}{(1-\alpha_{\ell_2})P+1}$. Upon successfully decoding the retransmitting packet, its interference can be removed using SIC, and the receiver can attempt to decode the fresh update with SNR $\gamma_2(\ell_2)={(1-\alpha_{\ell_2})P}$.
\end{IEEEproof}

The instantaneous MSE cost associated with state $J$ according to \eqref{eq:covarience_aoi} is $c_k(J_k,a_k) \triangleq \text {Tr}\left ({f^{q}(\bar {\mathbf {P}}_{0})}\right)$.  Under sufficient condition  \cite[Theorem 1]{huang2020real}, i.e., $\epsilon_\mathrm{cc}^{(m)}\rho^2({\mathbf{A}})<1$, the stationary and deterministic policy $\lambda_\mathrm{DN}$ exists that guarantees bounded long-term average MSE cost. The problem $\textrm{Prob.2a}$ is equivalent to the MDP average cost optimization problem. We use standard relative value iteration algorithms to solve this problem \cite{littman2013complexity}. 
Note that each combination of $\ell_1$ and $\ell_2$ in the MDP obtains a specific policy that gives the associated optimal average MSE cost defined in \eqref{eq:metric_average_MSE} and a specific value of cost variation given in \eqref{eq:metric_deviation} which falls less than the set $\theta$ limit in Prob.2a. In this way, the $L^2$ MDP's can give various policies leading to different cost function values that are Pareto optimal. Then the optimal policy can be selected that minimizes both objectives out of $L$ policies \cite{fan2016improved}.

\begin{remark}
Due to the high dimension state space involved, the computational complexity of the policy for DN-CC-HARQ is $\mathcal{O}(L^2N^{2}K)$ \cite{sennott2009stochastic}, where $N$ indicates the dimension of state space,  $K$ is the number of convergence steps. By reducing $L$, low-complexity solutions can be obtained with some loss in performance. Furthermore, by setting $\ell_2=\ell_1=\ell$ in the MDP,  a fixed (static) power-sharing fraction $\alpha_\ell$ based policy can be obtained with significantly less complexity. In practice, this case appears when the power-sharing fraction  $\alpha_\ell$ can be selected as $0\leq\alpha_\ell\leq1$ but remains fixed for all time slots.  
Then the complexity is reduces to $\mathcal{O}(LN^{2}K)$. We denote 
this policy as $\lambda_{\mathrm{SN}}(\alpha_\ell)$ as static N-CC-HARQ (SN-CC-HARQ) policy. The  Lemma 1 can be modified by setting $\ell_1=\ell_2=\ell$, which means $\alpha_{\ell_1}=\alpha_{\ell_2}=\alpha_{\ell}$, where $0\leq\alpha_{\ell}\leq 1$ to obtain the state transition function of MDP for SN-CC-HARQ. 
\end{remark}

\begin{remark}
The standard CC-HARQ-based policy optimization problem given in \cite[Eq (25)]{huang2020real} can be obtained by setting $\alpha_\ell=1$ in Lemma 1. This suggests that standard CC-HARQ is the sub-problem where only a single level of $\theta$ can be selected. Therefore, standard CC-HARQ gives a single point on the Pareto optimal front with complexity $\mathcal{O}(N^{2}K)$ \cite{sennott2009stochastic}. 
\end{remark}

\section{Numerical Results}
\label{Sec:Numarical_resutls}
 We use MATLAB-based MDP tool \cite{murphy2002markov} to solve each MDP problem for the optimal policy using the relative value iteration algorithm. We use the following simulation parameters unless specified otherwise: The LTI system dynamics are set as $\mathbf{A}=[2.4 , 0.2 ; 0.2, 0.8 ]$,
 $\mathbf{C}=[1,\; 1]$ $\mathbf{Q}_w=\mathbf{Q}_v=I$, $\rho^2(\mathbf{A})=1.8385^2$, $\bar{\mathbf{P}}_0=[2.5548,\; -1.6233;\quad -1.6233,\; 1.6719]$. $n_1=b=100$ for IR-HARQ and $n=n_1$ for CC-HARQ as the packet lengths during first transmission according to \eqref{eq:IR-HARQ_AW} and \eqref{eq:CC-HARQ_AW} respectively. All the simulations are conducted under AWGN channel conditions.
 
\subsection{Benchmark Schemes}
\textcolor{black}{
We first discuss various competing benchmark schemes as mentioned  in the introduction Section to highlight the importance of the proposed N-HARQ schemes. It is assumed that reducing the delay leads to MSE minimization that gives rise to AoI minimization design. However, with  HARQ, the situation changes because packet reliability can be increased with retransmission, which helps in minimizing the MSE by recovering the failing packets with a slight increase in delay due to retransmission. To obtain the  delay-optimal  policy for any given HARQ scheme under action $a\in\{0,1\}$, we change the cost function from MSE measure to AoI $q$. The policy for long-term average age minimization is obtained for solving the cost minimization problems  for different HARQ schemes by focusing on a single objective function as:}  
\begin{align} 
\lim _{K\to \infty }\frac {1}{K}\sum_{k=1}^{K} \mathbb{E}_w \left [{q_k}\right].
\label{eq:delay_optimal_simplef}
\end{align}
The optimal policies obtained through the above cost function are selected as the benchmark in \cite{huang2020real} and solved using a similar problem formulation as shown in Prob.1 and Prob.2 for IR-HARQ and CC-HARQ respectively with objective given in \eqref{eq:delay_optimal_simplef} and no constraint $\textrm{C}_1$. The MDP-based solution can be obtained as state space, action space, and state transition functions remain the same. However, the  cost of each state of the MDP of the delay-based policies is
\begin{align}
    c(m,q_k)|a_k=q_k.
\end{align}
The resulting policies are termed delay-based policies as they target to minimize the AoI. Apart from the delay-based policies, we compare other competing techniques such as ARQ-based \cite{gupta2010estimation}, fixed-HARQ-based, and optimized standard HARQ techniques \cite{huang2020real}.

\begin{figure}[t]
\centering
\includegraphics[width=0.95\columnwidth]{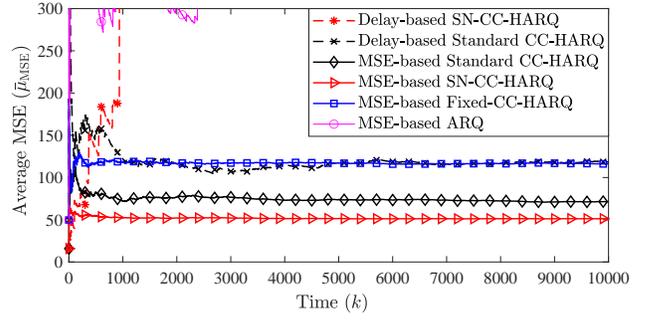}
\vspace{-0.1in}
\caption{Long-term average MSE performance comparison between  proposed SN-CC-HARQ ($\alpha_\ell=0.1$) and other benchmark  schemes at SNR=0dB, $n_1=b=100$ and $\rho^2(\mathbf{A})=2.4$.}
\label{fig:MSEvsAge}
\end{figure}
\textcolor{black}{In Fig. \ref{fig:MSEvsAge}, we provide the long-term average MSE performance comparison between MSE-based and delay-based policies for CC-HARQ. The Delay-based policy achieves poor MSE performance compared to the MSE-based policies with both standard CC-HARQ and proposed SN-CC-HARQ schemes. Because the MSE is a non-linear function of AoI \eqref{eq:covarience_aoi} and slight increases in AoI lead to much worst MSE. The higher performance loss of SN-CC-HARQ with the Delay-based policy is because excessive retransmission impacts the next packets carrying new status packets. As a result, we see in Fig. \ref{fig:MSEvsAge}, that Delay-based SN-CC-HARQ suffers much higher loss in comparison to standard CC-HARQ. Whereas  MSE-based policy gives the best result with SN-CC-HARQ. This clearly shows that in general, MSE cost functions are the best for designing policy optimization for applications involving real-time wireless remote estimation. It can be seen in Fig. \ref{fig:MSEvsAge} that the ARQ-based policies are worst mainly because it does not take advantage of increasing the reliability of status updates with retransmission. This clearly shows that the HARQ is required, but the fixed-HARQ scheme is not optimal because it does not care for the estimation MSE cost penalty with each retransmission. Also, note that the fixed-HARQ performs close to the Delay-based policy. Because fixed HARQ  does not adjust the reliability of the packet in relation to MSE cost penalty with retransmissions. MSE-based policies designed by \cite{huang2020real} take advantage of the trade-off between reliability and AoI to some extent but are limited due to the poor retransmission mechanism of HARQ. Next, we compare the  MSE-based policy due to \cite{huang2020real} and the proposed HARQ methods in detail.}

\subsection{IR-HARQ}
First, we see the IR-HARQ schemes and performance of optimal-policy obtained by solving MDP problem corresponding to  \eqref{eq:optimal_policy_IR_HARQ}.  The optimal policy $\lambda(\tau)$ achieves a particular MSE performance with a specific $\tau$. In IR-HARQ, we find optimal policy under optimized retransmission parameter $\tau$ to achieve the best MSE performance.   

\begin{figure}[t]
\centering
\includegraphics[width=0.95\columnwidth]{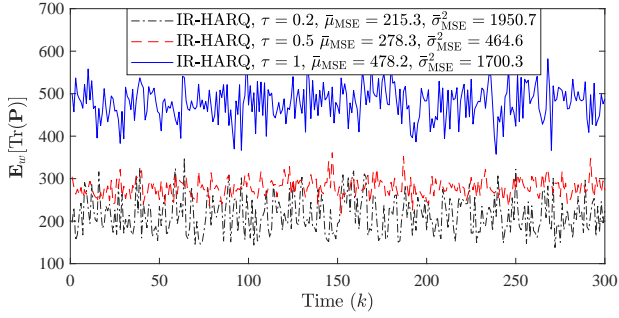}
\vspace{-0.1in}
\caption{\textcolor{black}{ MSE performance variation of IR-HARQ with varying $\tau$, $n_1=b=100$ at SNR=0dB, $m=2$ and $\rho^2(\mathbf{A})=4.4$.} }
\label{fig:MSEvsAge_variatin}
\end{figure}
\textcolor{black}{In Fig. \ref{fig:MSEvsAge_variatin}, we show the performance of MSE-policy obtained due to the IR-HARQ scheme at three different $\tau$ values. It can be seen that under specific policy $\lambda(\tau)$, the instantaneous MSE $\mathrm{Tr}(\mathbf{P}_k)$ varies over each time slot due to packet success or failure. Therefore, we perform 1000  Monte Carlo rounds to characterize the expected MSE performance ($\mathbb{E}_w$) over $K=300$ time slots with a specific policy, where the random variable $w$ indicates the packet errors due to short packet duration and AWGN channel. This means that the expectation $\mathbb{E}_w$ is taken over  1000 instances. The average MSE and its variation can be calculated according to \eqref{eq:metric_average_MSE} and \eqref{eq:metric_deviation}, respectively. Finally, we can see that the optimal policy with $\tau=0.5$ gives the most stable MSE performance and relatively better average MSE over $K=300$ time slots. }

\begin{figure}[t]
\centering
\includegraphics[width=0.95\columnwidth]{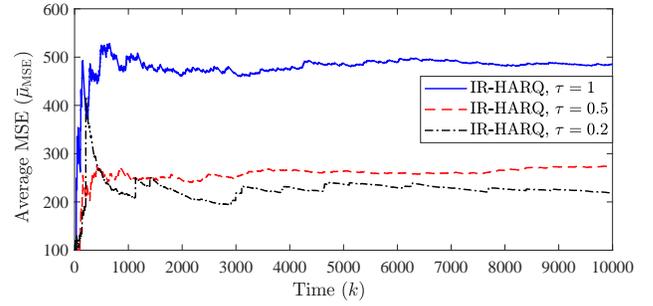}
\vspace{-0.1in}
\caption{Average MSE performance of IR-HARQ due to various  $\tau$ at SNR=0dB when $n_1=b=100$ and $\rho^2(\mathbf{A})=4.4$. }
\label{fig:MSE_IR_HARQ}
\end{figure}

Fig. \ref{fig:MSE_IR_HARQ}  presents the impact of $\tau$ on the long-term average MSE performance of IR-HARQ. Fig. \ref{fig:MSE_IR_HARQ} shows that the smaller values of $\tau$, e.g., $0.2$ and $0.5$, lead to a lower average  MSE. The IR-HARQ policy of \cite{huang2020real} at $\tau=1$ is sub-optimal as it sends the full retransmission and increases the AoI to the maximum with each retransmission. The age penalty due to retransmitting packets reduces with smaller $\tau$. Fig. \ref{fig:MSE_IR_HARQ} indicates that the performance gain is higher when reducing $\tau$ from  1 to 0.5 than from 0.5 to 0.2. As $\tau$ reduces, the packet reliability also reduces according to \eqref{eq:IR-HARQ_AW}. Therefore, smaller $\tau$ may cause consecutive packet failures in many time slots, especially when fewer retransmissions are allowed. In this situation, the AoI $q$ and MSE grow much faster and more often, as seen in higher fluctuations when $\tau=0.2$ in Fig. \ref{fig:MSE_IR_HARQ}. Therefore, the optimal $\tau$ that minimizes the standard deviation may not always be the lowest $\tau$. 

\begin{figure}[t]
\centering
\includegraphics[width=0.95\columnwidth]{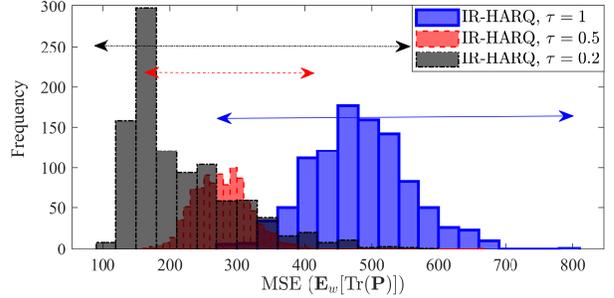}
\vspace{-0.1in}
\caption{MSE distribution of IR-HARQ  due to various $\tau$ at SNR=0dB and $n_1=b=100$ and $\rho^2(\mathbf{A})=4.4$.}
\label{fig:MSE_hist}
\end{figure}
In real-time remote estimation, the variance of MSE during policy execution is also important. In Fig. \ref{fig:MSE_hist}, we plot the distribution of the MSE performance  under the optimal policy at various $\tau$ settings. We test the performance of optimal policy by sending packets  and drawing the MSE distribution using the Monte Carlo simulations for $K=1000$ time slots. The MSE distribution shows the relative frequency of MSE fluctuations at various $\tau$ settings. As can be seen in Fig. \ref{fig:MSE_hist}, the deviation of MSE from the mean is higher with $\tau=0.2$ and $\tau=1$ than $\tau=0.5$. In real-time remote estimation, if a packet fails, the receiver estimates the current state using the last successful update with higher AoI $q$. At $\tau=0.2$, HARQ retransmissions provide very low reliability to failing packets, and erroneous packets are mostly not recovered after the retransmission. Therefore,  MSE deviates significantly from the mean with higher frequency in different time slots. Whereas the setting $\tau=1$ provides excessive reliability at the cost of higher AoI in each time slot due to retransmission. Specifically, each retransmission at $\tau=1$ causes a higher MSE  penalty leading to an overall higher MSE deviation.  $\tau=0.5$ provides better reliability with the lowest possible AoI growth with each retransmission. Finally, the  policy with $\tau=0.5$ would be selected to get optimal performance for the optimization problem \eqref{eq:optimal_policy_IR_HARQ}. 
\vspace{-0.1in}
\subsection{CC-HARQ}
\begin{figure}[t]
\centering
\includegraphics[width=0.95\columnwidth]{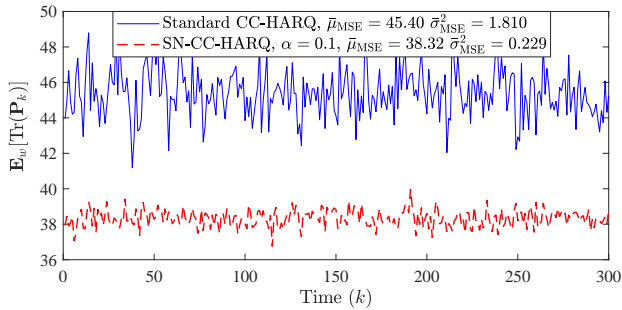}
\vspace{-0.1in}
\caption{\textcolor{black}{Impact of optimal policy with SN-CC-HARQ and standard CC-HARQ over estimation MSE variance at SNR=0dB and $n_1=b=100$ and $\rho^2(\mathbf{A})=2.0$.} }
\label{fig:CC_variance}
\end{figure}
\begin{figure}[t]
\centering
\includegraphics[width=0.95\columnwidth]{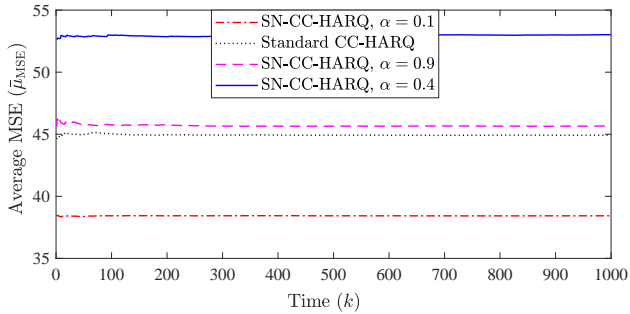}
\vspace{-0.1in}
\caption{Average MSE performance of SN-CC-HARQ  due to various $\alpha$ at SNR=0dB and $n_1=b=100$ and $\rho^2(\mathbf{A})=2.0$.}
\label{fig:N_CC_MSE}
\end{figure}

\textcolor{black}{
In Fig. \ref{fig:CC_variance}, we show the variation of MSE performance with SN-CC-HARQ-based policy at $\alpha=0.1$ and standard CC-HARQ-based optimal policy. We use 1000 trails of packet transmission under a specific policy to characterize the average MSE variation with each packet transmission to show its variance over 300 time slots. It can be seen that the proposed SN-CC-HARQ method achieves a better average MSE performance with low variation. Next, we elaborate further on the reasons for the performance improvements with more results and comparisons.}

In Fig. \ref{fig:N_CC_MSE}, we compare the MSE performance of standard CC-HARQ with SN-CC-HARQ. The standard CC-HARQ retransmits the old update with maximum allocated time slot. In contrast, SN-CC-HARQ conducts retransmission using power-sharing parameter $\alpha$ to update current status along with retransmission of old update. We show the effect of selecting different $\alpha$ on the long-term average MSE performance of SN-CC-HARQ in Fig. \ref{fig:N_CC_MSE}. As can be seen in  this figure,  with SN-CC-HARQ, $\alpha=0.1$  achieves the best long-term average MSE performance followed by the setting  $\alpha=0.9$, while  $\alpha=0.4$ gives the worst performance. 
This is because at setting $\alpha=0.1$ and $0.9$, the power difference between non-orthogonal pocket is higher which leads to better decoding under SIC. When the overlapping packets are at similar power levels, e.g., $\alpha=0.4$, it is challenging to separate non-orthogonal packets successfully using SIC. Comparing $\alpha=0.1$ and $\alpha=0.9$, the setting $\alpha=0.9$ assigns excessive power for sending old updates, which leaves less power for fresh status updates and causes estimation errors. The standard CC-HARQ also suffers from poor MSE performance due to excessive retransmission overhead. Since at $\rho^2(\mathbf{A})=2.0$, the correlation between status updates is higher and fresh updates with a little higher reliability is more suitable. At $\alpha=0.1$, the non-orthogonal fresh updates are  decoded with good reliability while utilizing enough retransmissions to save failing old updates. Therefore, under given packet reliability and process correlation profile, i.e., $\rho^2(\mathbf{A})=2.0$, the optimal policy with $\alpha=0.1$ achieves the best long-term average MSE performance.

\begin{figure}[t]
\centering
\includegraphics[width=0.95\columnwidth]{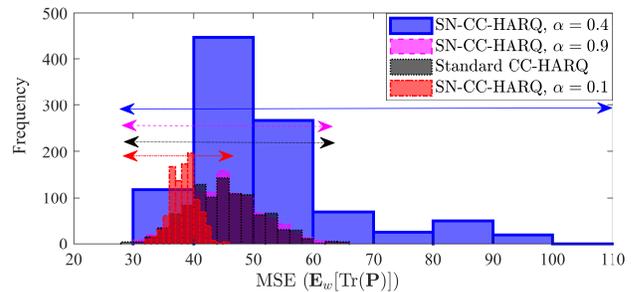}
\vspace{-0.1in}
\caption{MSE distribution of SN-CC-HARQ  due to various $\alpha$  at SNR=0dB when $n_1=b=100$ and $\rho^2(\mathbf{A})=2.0$.}
\label{fig:N_CC_Hist}
\end{figure}

The  histogram of  MSE in Fig. \ref{fig:N_CC_Hist} depicts the variation of MSE given  due to SN-CC-HARQ-based policy $\lambda_\mathrm{SN}(\alpha)$ corresponding to  various $\alpha$ values. In Fig.\ref{fig:N_CC_Hist}, we show the MSE performance comparison between standard CC-HARQ and SN-CC-HARQ. At $\alpha=0.4$, the policy leads to unstable MSE performance, primarily due to inaccurate SIC decoding of SN-CC-HARQ. The packet failure rate is reduced by increasing the retransmission power from $\alpha=0.4$ to $\alpha=0.9$. However, at $\alpha=0.9$, the retransmission power is excessive, and the overlapping fresh update is transmitted with less power ($1-\alpha=0.1$) which would require a retransmission. Excessive retransmissions over many time slots increase the AoI and lead to sub-optimal and unstable MSE performance, as can be seen in Fig. \ref{fig:N_CC_Hist}, for $\alpha=0.9$. The MSE variation of standard CC-HARQ is almost similar to SN-CC-HARQ at $\alpha=0.9$ because of roughly the same power assignment to send old updates considering the correlation of the dynamic process $\rho^2(\mathbf{A})=2.0$. At $\alpha=0.1$ sensor uses the exact required power to conduct retransmission leaving enough power to send fresh update. This results in more packets being decoded successfully in consecutive time slots leading to better and more stable MSE performance.

\begin{figure}[tb]
\centering
\includegraphics[width=0.95\columnwidth]{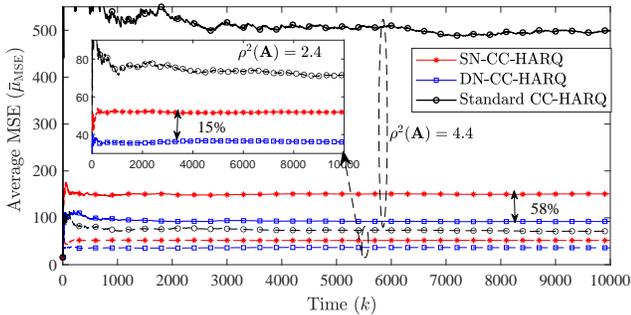}
\vspace{-0.1in}
\caption{Average MSE performance of SN-CC-HARQ with  $\alpha=0.1$  and DN-CC-HARQ with $L=11$ power levels, i.e., $\alpha_{\ell}\in\{0,0.1,0.2,\cdots 1\}$ in AWGN channel at $n_1=b=100$, SNR=0dB and $m=2$.}
\label{fig:MSE_dynamic-static}
\end{figure}

Fig. \ref{fig:MSE_dynamic-static} shows a performance comparison between proposed CC-HARQ schemes and standard CC-HARQ schemes due to increasing $\rho^2(\mathbf{A})$ values.
The eigenvalue $\rho^2(\mathbf{A})$ of the LTI system matrix represents the correlation of the dynamic process. In Fig. \ref{fig:MSE_dynamic-static} the solid lines corresponds to setting $\rho^2(\mathbf{A})=4.4$ and bottom 3 dotted lines show performance at $\rho^2(\mathbf{A})=2.4$. The gap between the proposed scheme and baseline standard CC-HARQ increases significantly by increasing  $\rho^2(\mathbf{A})$ from 2.4 to 4.4. Also,  Fig. \ref{fig:MSE_dynamic-static}  shows that SN-CC-HARQ and DN-CC-HARQ schemes are more effective in limiting MSE cost penalty due to higher flexibility in selecting reliability and AoI setting. The worst-performing Standard CC-HARQ takes a retransmission action with full power and time slot utilization, which increases the AoI, causing severe MSE loss, especially  when the correlation is low.

 Under the dynamic policy, the sensor can vary $\alpha_\ell$ in each time slot leading to further performance improvements, as evident in Fig. \ref{fig:MSE_dynamic-static}. For example, suppose the sensor allocates higher $\alpha$ for retransmission of old update at time slot $k-1$. It can reduce its impact by allocating more retransmission power for the retransmission of the following status update; leading to more controlled growth of AoI. More specifically, when the AoI is low, the sensor chooses a smaller value of $\alpha_\ell$, i.e.,  $\{0, 0.1, \cdots,0.5\}$ for AoI  $\{1, 2, \cdots,6\}$ respectively. When AoI is higher due to consecutive packet failure, the sensor selects higher power levels, such as $0.9$ and $1$, to limit the AoI growth. DN-CC-HARQ gives 15$\%$ performance improvements over the SN-CC-HARQ when $\rho^2(\mathbf{A})=2.4$. Furthermore, when the correlation between status updates is low, i.e., $\rho^2(\mathbf{A})=4.4$, the MSE gain of dynamic policy over static is increased with a 58$\%$ gain. Each power fraction $\alpha_\ell$ corresponds to a specific packet reliability versus AoI setting. Dynamically selecting $\alpha$ gives the sensor the flexibility to choose suitable reliability during retransmission in each time slot. On the other hand, in SN-CC-HARQ, the choice of $\alpha$ remains fixed and associated with specific reliability versus AoI growth with less complexity. Yet it performs significantly better than the existing standard CC-HARQ.

\begin{figure}[t]
\centering
\includegraphics[width=0.95\columnwidth]{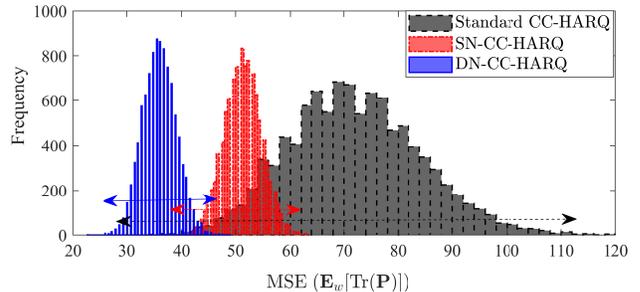}
\vspace{-0.1in}
\caption{MSE distributions of standard CC-HARQ, proposed SN-CC-HARQ with  $\alpha=0.1$  and DN-CC-HARQ with $L=11$ power levels, i.e., $\alpha_\ell\in\{0,0.1,0.2,\cdots 1\}$ in AWGN channel at $n_1=b=100$, SNR=0dB and $m=2$.}
\label{fig:MSE_hist_CC}
\end{figure}
Fig. \ref{fig:MSE_hist_CC} shows the histograms of the  MSE for various CC-HARQ schemes under corresponding optimal policies.  The performance of optimal policy with standard  CC-HARQ performs worst in providing stable and low MSE. This is because standard CC-HARQ increases packet reliability with maximum increases in AoI, leading to poor and highly unstable MSE performance. The proposed N-CC-HARQ can avoid that by adjusting the AoI growth using an additional variable $\alpha$. The  SN-CC-HARQ reduces the MSE and provides more controlled MSE growth in each time slot. The DN-CC-HARQ-based policy can further limit the MSE deviation compared to the SN-CC-HARQ-based policy. This is due to higher flexibility and control over AoI growth with retransmissions in each time slot. 
\section{Practical Considerations} 
\label{Sec:Practical_Considerations}
In the non-orthogonal CC-HARQ, the packets are required to be separated first using SIC,  which adds to the decoding complexity. Similarly, due to multipacket decoding, a multi-bit feedback mechanism would be required. At this stage, we rely on the NA to characterize the performance of the proposed schemes in relation to standard HARQ. Therefore, we assume perfect feedback and synchronization, etc. However, the impact of these assumptions would become more relevant when using practical encoder decoders. For IR-HARQ, due to partial retransmission ($\tau$), there involves some synchronization overhead. 

In ARQ and fixe-HARQ schemes, the receiver can always differentiate between retransmission and new packet. However, the most important implementation challenge for policy-based design, including the baseline \cite{huang2020real}, is the synchronization of the policy to identify the transmission and new packet for packet combining. The first solution is to share the policy between the transmitter and receiver. Note that the policy is the mapping of states to actions, and we use MDP with a finite number of states due to ($q_{max}=10$, $m=2$ ). Then transmitter determines the latest state (because state change is a stochastic due erroneous channel) of the receiver through feedback to determine the next action. This solution is inferior due to an overhead of extra bits whenever the policy changes. Often the one-to-one mapping of the state to actions can be transformed into simpler structures that can reduce the overhead for practical purposes \cite{blok2017structures}. The other solution is to use extra bits in each packet to differentiate between a retransmission packet and new transmission. But this solution also comes with a few bits but more frequent overhead than policy synchronization. Fortunately, the MDP problems give many possibilities for prediction of the hidden Markov structure that can be used to form a consensus between transmitter and receiver to learn policy online gradually \cite{hefny2018efficient}. However, in this work, we focus more on the design of modified HARQ methods.

\section{Conclusion}
We proposed wireless network control for remote estimation of the LTI dynamical systems under various IR-HARQ  and CC-HARQ-based packet retransmission schemes. For IR-HARQ, optimized retransmission improved the MSE performance due to the early arrival of fresh status updates. We optimized the standard CC-HARQ schemes and allowed non-orthogonal retransmissions that increased the packet reliability gradually and limited the AoI and MSE growth. The power-sharing fraction $\alpha$ is optimized for the sensor policy under the non-orthogonal CC-HARQ scheme. In dynamic-optimal policy, $\alpha$ is allowed to vary for greater flexibility, which results in significant performance gain with higher complexity. In the static-optimal policy, $\alpha$ remained fixed to reduce the complexity but still achieved much better MSE performance compared to the optimal policy with standard CC-HARQ. We used the Markov decision process formulation to solve complex bi-objective optimization problems and obtain optimal policies using numerical techniques. Simulation results exhibited a significantly better long-term average MSE performance as well as low MSE variance in each time slot that guaranteed better performance stability for the time-sensitive applications.
\label{sec:conclusion}
  
\footnotesize
\bibliographystyle{IEEEtran}
\bibliography{myReferences}

\begin{thebibliography}{10}
\providecommand{\url}[1]{#1}
\csname url@samestyle\endcsname
\providecommand{\newblock}{\relax}
\providecommand{\bibinfo}[2]{#2}
\providecommand{\BIBentrySTDinterwordspacing}{\spaceskip=0pt\relax}
\providecommand{\BIBentryALTinterwordstretchfactor}{4}
\providecommand{\BIBentryALTinterwordspacing}{\spaceskip=\fontdimen2\font plus
\BIBentryALTinterwordstretchfactor\fontdimen3\font minus
  \fontdimen4\font\relax}
\providecommand{\BIBforeignlanguage}[2]{{%
\expandafter\ifx\csname l@#1\endcsname\relax
\typeout{** WARNING: IEEEtran.bst: No hyphenation pattern has been}%
\typeout{** loaded for the language `#1'. Using the pattern for}%
\typeout{** the default language instead.}%
\else
\language=\csname l@#1\endcsname
\fi
#2}}
\providecommand{\BIBdecl}{\relax}
\BIBdecl

\bibitem{varrall20165g}
G.~Varrall, \emph{5G spectrum and standards}.\hskip 1em plus 0.5em minus
  0.4em\relax Artech House, 2016.

\bibitem{antonakoglou2018toward}
K.~Antonakoglou, X.~Xu, E.~Steinbach, T.~Mahmoodi, and M.~Dohler, ``Toward
  haptic communications over the {5G} tactile {I}nternet,'' \emph{IEEE
  Communications Surveys \& Tutorials}, vol.~20, no.~4, pp. 3034--3059, 2018.

\bibitem{peng2013event}
C.~Peng and T.~C. Yang, ``Event-triggered communication and control co-design
  for networked control systems,'' \emph{Automatica}, vol.~49, no.~5, pp.
  1326--1332, 2013.

\bibitem{peng2021sensing}
F.~Peng, Z.~Jiang, S.~Zhou, Z.~Niu, and S.~Zhang, ``Sensing and communication
  co-design for status update in multiaccess wireless networks,'' \emph{IEEE
  Transactions on Mobile Computing}, 2021.

\bibitem{bennis2018ultrareliable}
M.~Bennis, M.~Debbah, and H.~V. Poor, ``Ultrareliable and low-latency wireless
  communication: Tail, risk, and scale,'' \emph{Proceedings of the IEEE}, vol.
  106, no.~10, pp. 1834--1853, 2018.

\bibitem{yates2021age}
R.~D. Yates, Y.~Sun, D.~R. Brown, S.~K. Kaul, E.~Modiano, and S.~Ulukus, ``Age
  of information: An introduction and survey,'' \emph{IEEE Journal on Selected
  Areas in Communications}, vol.~39, no.~5, pp. 1183--1210, 2021.

\bibitem{gupta2010estimation}
V.~Gupta, ``On estimation across analog erasure links with and without
  acknowledgements,'' \emph{IEEE transactions on automatic control}, vol.~55,
  no.~12, pp. 2896--2901, 2010.

\bibitem{ornee2021sampling}
T.~Z. Ornee and Y.~Sun, ``Sampling and remote estimation for the
  ornstein-uhlenbeck process through queues: Age of information and beyond,''
  \emph{IEEE/ACM Transactions on Networking}, 2021.

\bibitem{roth2020remote}
S.~Roth, A.~Arafa, H.~V. Poor, and A.~Sezgin, ``Remote short blocklength
  process monitoring: Trade-off between resolution and data freshness,'' in
  \emph{ICC 2020-2020 IEEE International Conference on Communications
  (ICC)}.\hskip 1em plus 0.5em minus 0.4em\relax IEEE, 2020, pp. 1--6.

\bibitem{Rajaraman2021NotJA}
N.~Rajaraman, R.~Vaze, and G.~Reddy, ``Not just age but age and quality of
  information,'' \emph{IEEE Journal on Selected Areas in Communications},
  vol.~39, pp. 1325--1338, 2021.

\bibitem{wang2020value}
Z.~Wang, M.-A. Badiu, and J.~P. Coon, ``A value of information framework for
  latent variable models,'' in \emph{GLOBECOM 2020-2020 IEEE Global
  Communications Conference}.\hskip 1em plus 0.5em minus 0.4em\relax IEEE,
  2020, pp. 1--6.

\bibitem{huang2020real}
K.~Huang, W.~Liu, M.~Shirvanimoghaddam, Y.~Li, and B.~Vucetic, ``Real-time
  remote estimation with hybrid {ARQ} in wireless networked control,''
  \emph{IEEE Transactions on Wireless Communications}, vol.~19, no.~5, pp.
  3490--3504, 2020.

\bibitem{zhang2020energy}
B.~Zhang, L.~B. Milstein, and P.~Cosman, ``Energy optimization for hybrid {ARQ}
  with turbo coding: Rate adaptation and allocation,'' \emph{IEEE Transactions
  on Vehicular Technology}, vol.~69, no.~10, pp. 11\,338--11\,352, 2020.

\bibitem{nadeem2021analysis}
F.~Nadeem, Y.~Li, B.~Vucetic, and M.~Shirvanimoghaddam, ``Analysis and
  optimization of {HARQ} for {URLLC},'' in \emph{2021 IEEE Globecom Workshops
  (GC Wkshps)}.\hskip 1em plus 0.5em minus 0.4em\relax IEEE, 2021, pp. 1--6.

\bibitem{faisalMDPI}
F.~Nadeem, M.~Shirvanimoghaddam, Y.~Li, and B.~Vucetic, ``Delay-sensitive
  {NOMA-HARQ} for short packet communications,'' \emph{Entropy}, vol.~1, no.~0,
  pp. 000--000, 2021.

\bibitem{nadeem2022real}
F.~Nadeem, Y.~Li, B.~Vucetic, and M.~Shirvanimoghaddam, ``Real-time wireless
  control with non-orthogonal {HARQ},'' in \emph{2022 IEEE Globecom Workshops
  (GC Wkshps)}.\hskip 1em plus 0.5em minus 0.4em\relax IEEE, 2022, pp. 1--6.

\bibitem{schenato2008optimal}
L.~Schenato, ``Optimal estimation in networked control systems subject to
  random delay and packet drop,'' \emph{IEEE transactions on automatic
  control}, vol.~53, no.~5, pp. 1311--1317, 2008.

\bibitem{liu2016energy}
W.~Liu, X.~Zhou, S.~Durrani, H.~Mehrpouyan, and S.~D. Blostein, ``Energy
  harvesting wireless sensor networks: Delay analysis considering energy costs
  of sensing and transmission,'' \emph{IEEE Transactions on Wireless
  Communications}, vol.~15, no.~7, pp. 4635--4650, 2016.

\bibitem{maybeck1982stochastic}
P.~S. Maybeck, \emph{Stochastic models, estimation, and control}.\hskip 1em
  plus 0.5em minus 0.4em\relax Academic press, 1982.

\bibitem{rhudy2013online}
M.~B. Rhudy and Y.~Gu, ``Online stochastic convergence analysis of the kalman
  filter.'' \emph{International Journal of Stochastic Analysis}, 2013.

\bibitem{erseghe2016coding}
T.~Erseghe, ``Coding in the finite-blocklength regime: Bounds based on laplace
  integrals and their asymptotic approximations,'' \emph{IEEE Transactions on
  Information Theory}, vol.~62, no.~12, pp. 6854--6883, 2016.

\bibitem{nadeem2021non}
F.~Nadeem, M.~Shirvanimoghaddam, Y.~Li, and B.~Vucetic, ``Non-orthogonal {HARQ}
  for {URLLC}: Design and analysis,'' \emph{IEEE Internet of Things Journal},
  2021.

\bibitem{polyanskiy2010channel}
Y.~Polyanskiy, H.~V. Poor, and S.~Verd{\'u}, ``Channel coding rate in the
  finite blocklength regime,'' \emph{IEEE Transactions on Information Theory},
  vol.~56, no.~5, p. 2307, 2010.

\bibitem{sahin2019delay}
C.~Sahin, L.~Liu, E.~Perrins, and L.~Ma, ``Delay-sensitive communications over
  {IR-HARQ}: Modulation, coding latency, and reliability,'' \emph{IEEE Journal
  on Selected Areas in Communications}, vol.~37, no.~4, pp. 749--764, 2019.

\bibitem{kaul2012real}
S.~Kaul, R.~Yates, and M.~Gruteser, ``Real-time status: How often should one
  update?'' in \emph{2012 Proceedings IEEE INFOCOM}.\hskip 1em plus 0.5em minus
  0.4em\relax IEEE, 2012, pp. 2731--2735.

\bibitem{shi2012scheduling}
L.~Shi and H.~Zhang, ``Scheduling two {Gauss--Markov} systems: An optimal
  solution for remote state estimation under bandwidth constraint,'' \emph{IEEE
  Transactions on Signal Processing}, vol.~60, no.~4, pp. 2038--2042, 2012.

\bibitem{wu2020optimal}
S.~Wu, K.~Ding, P.~Cheng, and L.~Shi, ``Optimal scheduling of multiple sensors
  over lossy and bandwidth limited channels,'' \emph{IEEE Transactions on
  Control of Network Systems}, vol.~7, no.~3, pp. 1188--1200, 2020.

\bibitem{ashok2017value}
P.~Ashok, K.~Chatterjee, P.~Daca, J.~K{\v{r}}et{\'\i}nsk{\`y}, and
  T.~Meggendorfer, ``Value iteration for long-run average reward in markov
  decision processes,'' in \emph{International Conference on Computer Aided
  Verification}.\hskip 1em plus 0.5em minus 0.4em\relax Springer, 2017, pp.
  201--221.

\bibitem{branke2008multiobjective}
J.~Branke, J.~Branke, K.~Deb, K.~Miettinen, and R.~Slowi{\'n}ski,
  \emph{Multiobjective optimization: Interactive and evolutionary
  approaches}.\hskip 1em plus 0.5em minus 0.4em\relax Springer Science \&
  Business Media, 2008, vol. 5252.

\bibitem{haimes1971bicriterion}
Y.~Haimes, ``On a bicriterion formulation of the problems of integrated system
  identification and system optimization,'' \emph{IEEE transactions on systems,
  man, and cybernetics}, vol.~1, no.~3, pp. 296--297, 1971.

\bibitem{chiandussi2012comparison}
G.~Chiandussi, M.~Codegone, S.~Ferrero, and F.~E. Varesio, ``Comparison of
  multi-objective optimization methodologies for engineering applications,''
  \emph{Computers \& Mathematics with Applications}, vol.~63, no.~5, pp.
  912--942, 2012.

\bibitem{littman2013complexity}
M.~L. Littman, T.~L. Dean, and L.~P. Kaelbling, ``On the complexity of solving
  markov decision problems,'' \emph{arXiv preprint arXiv:1302.4971}, 2013.

\bibitem{fan2016improved}
Z.~Fan, H.~Li, C.~Wei, W.~Li, H.~Huang, X.~Cai, and Z.~Cai, ``An improved
  epsilon constraint handling method embedded in moea/d for constrained
  multi-objective optimization problems,'' in \emph{2016 IEEE Symposium Series
  on Computational Intelligence (SSCI)}.\hskip 1em plus 0.5em minus 0.4em\relax
  IEEE, 2016, pp. 1--8.

\bibitem{sennott2009stochastic}
L.~I. Sennott, \emph{Stochastic dynamic programming and the control of queueing
  systems}.\hskip 1em plus 0.5em minus 0.4em\relax John Wiley \& Sons, 2009,
  vol. 504.

\bibitem{murphy2002markov}
M.-J. Cros, ``Markov decision process ({MDP}) toolbox for {MATLAB},''
  \emph{MATLAB Central file exchange, [Online] Available:
  {(https://au.mathworks.com/matlabcentral/fileexchange/25786-markov-decision-processes-mdp-toolbox)}},
  2002.

\bibitem{blok2017structures}
H.~Blok and F.~Spieksma, ``Structures of optimal policies in {MDP}s with
  unbounded jumps: the state of our art,'' in \emph{Markov Decision Processes
  in Practice}.\hskip 1em plus 0.5em minus 0.4em\relax Springer, 2017, pp.
  131--186.

\bibitem{hefny2018efficient}
A.~Hefny, ``Efficient methods for prediction and control in partially
  observable environments,'' Ph.D. dissertation, Carnegie Mellon University,
  2018.

\end{thebibliography}

\end{document}